\documentclass[%
reprint,
superscriptaddress,
preprintnumbers,
 amsmath,amssymb,
 aps,
prb,
floatfix,
]{revtex4-2}

\usepackage{xcolor}
\usepackage[normalem]{ulem}
\usepackage{color,soul} 
\usepackage{amsmath}

\usepackage{graphicx}
\usepackage{dcolumn}
\usepackage{bm}
\usepackage[utf8]{inputenc}

\begin{document}

\title{Observation of Electrically Tunable Chirality Inversion in a Slow-Light Waveguide}

\author{Xuchao Chen}
\author{Savvas Germanis}
 \email{s.germanis@sheffield.ac.uk}
\affiliation{School of Mathematical and Physical Sciences, University of Sheffield, Hicks Building, Sheffield S3 7RH, United Kingdom }
\author{Nicholas J. Martin}
 \email{n.j.martin@sheffield.ac.uk}
\affiliation{School of Mathematical and Physical Sciences, University of Sheffield, Hicks Building, Sheffield S3 7RH, United Kingdom }
\author{Hamidreza Siampour}
\affiliation{School of Mathematics and Physics, Queen's University Belfast, University Road, Belfast BT7 1NN, United Kingdom}
\author{Ren\'e Dost}
\author{Dominic J. Hallett}
\affiliation{School of Mathematical and Physical Sciences, University of Sheffield, Hicks Building, Sheffield S3 7RH, United Kingdom }
\author{Ian Farrer}
\author{Akshay Kumar Verma}
\affiliation{School of Electrical and Electronic Engineering, University of Sheffield, Sheffield S1 3JD, United Kingdom}
\author{Maurice S. Skolnick}
\author{Luke R. Wilson}
\author{A. Mark Fox}
\affiliation{School of Mathematical and Physical Sciences, University of Sheffield, Hicks Building, Sheffield S3 7RH, United Kingdom }

            
\begin{abstract}
We identify chiral inversion points in slow-light, glide-plane-symmetric, photonic-crystal waveguides, defined as fixed locations where the local optical chirality changes sign over a narrow wavelength range. We experimentally access this behaviour using a waveguide-embedded InAs/InGaAs quantum dot. The slow-light spectral region is determined from time-integrated and time-resolved photoluminescence, and the dot exciton is electrically tuned across the slow-light bandwidth via the quantum-confined Stark effect. As the emission wavelength is swept through the slow-light region, the directional emission contrast shows a strong wavelength dependence and a sign reversal, consistent with the identified chiral inversion point. Numerical simulations attribute the switching primarily to the pronounced spectral variation of the local optical chirality for emitters displaced from the waveguide center. These results demonstrate on-demand electrical switching of chiral light-matter coupling in nanophotonic waveguides and enable tunable chiral interfaces for integrated quantum photonic devices.
\end{abstract}

\maketitle

\section{Introduction}

\begin{figure*}[t]
\includegraphics[width=\textwidth]{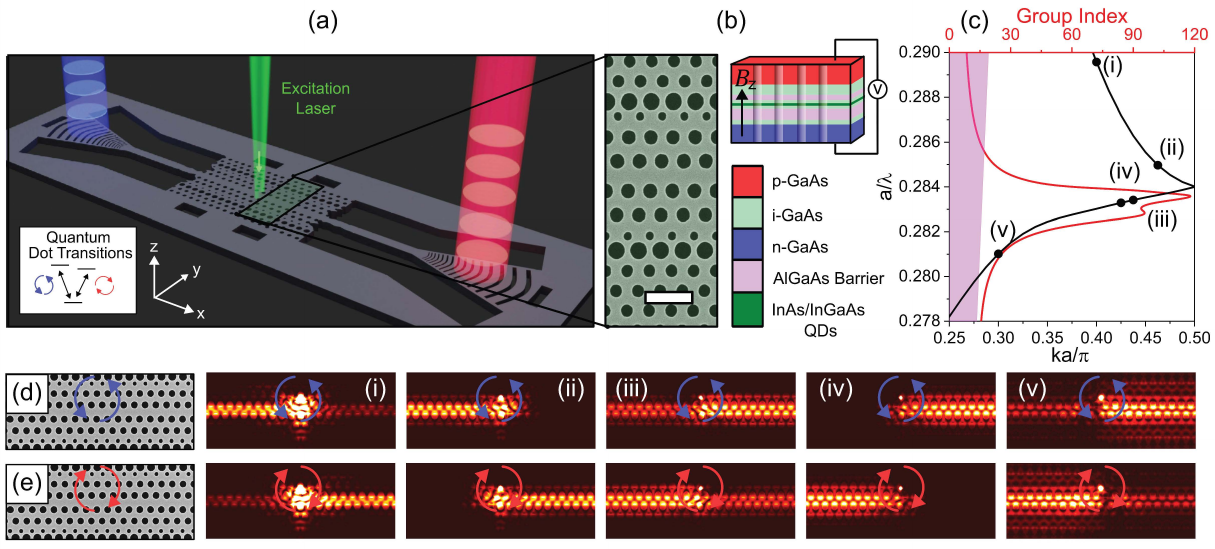}
\caption{(a) Device schematic layout and working principle. A QD located in the active region of the waveguide is optically excited, and the emitted photons are guided toward grating out-couplers at both ends for photoluminescence collection. (b) An SEM image showing a part of the active region of GPW device, together with a schematic of the p-i-n diode structure. The scale bar represents 500~nm. (c) Simulated band structure of the central GPW region and the extracted group index $n_g$. Five sampling wavelengths are highlighted on the dispersion curve in an increasing order and are used for (d)-(e). The light cone is shaded in purple. (d-e) Schematic of the directional coupling of a left/right circularly polarised dipole into the waveguide mode. The dipole is placed at an off-centred location, aligned horizontally with the centre of the second innermost air-hole and vertically displaced by $1.3a$ from the centre of the waveguide. Panel (i-v) show the coupling behaviour at five selected wavelengths (i.e. $a/\lambda=0.2810, 0.2833, 0.2834, 0.2838, 0.2895$).  }
\label{fig1}
\end{figure*}

Waveguide quantum electrodynamics (QED) provides a versatile platform for controlling the interaction between single quantum emitters and guided optical modes, and is therefore a key ingredient for integrated quantum photonic technologies. By coupling solid-state emitters to one-dimensional nanophotonic waveguides, spontaneous emission can be directed efficiently into propagating modes, enabling on-chip generation, routing, and manipulation of single photons \cite{Shen2005, Chang2014, Lodahl2015, Nieddu2016, Bhaskar2017, Turschmann2019, Prasad2020, Blais2021, Kim2021}. Semiconductor quantum dots (QDs) are particularly attractive in this context because they combine high brightness and spectral stability with compatibility with established nanofabrication methods \cite{Lund-Hansen2008, Makhonin2014,Coles2017}. Gallium arsenide platforms further offer a versatile route toward monolithic integration of emitters, passive circuitry, and active functionalities \cite{Dietrich2016}.

In nanophotonic waveguides, the interaction between a quantum emitter, such as a QD, and a guided optical mode can depend on both the propagation direction of the mode and the polarisation of the optical transition. Because the local handedness of the guided field reverses when the propagation direction is reversed, the emission direction is determined by the overlap between the handedness of the quantum-dot transition and the local mode polarisation. This leads to chiral light-matter interactions, in which emission into the left- and right-propagating modes becomes asymmetric. Such direction-dependent coupling has been demonstrated in a wide range of nanophotonic systems, including photonic-crystal and nanobeam waveguides \cite{Mahmoodian2016,Coles2017,Sollner2015,Javadi2018}, optical nanofibres \cite{Petersen2014,Mitsch2014}, topological waveguides \cite{Martin_comparison}, cavity-based platforms \cite{Martin2025,Hallacy2025TopologicalInterface,Antoniadis2022}, and ring-resonator geometries \cite{Mehrabad2020,Mehrabad2023AddDrop,Barik2020,Rao2025}. This phenomenon is a manifestation of spin-momentum locking of light \cite{Bliokh2015,Lodahl2017,PRXQuantum.6.020101}, and is of considerable interest for directional single-photon sources, spin-photon interfaces, and nonreciprocal quantum optical elements \cite{Sollner2015,Coles2016,Javadi2018,PRXQuantum.6.020101,switch_review}.

Among the available platforms, glide-plane photonic-crystal waveguides are particularly well suited to chiral quantum optics because their symmetry properties support circularly polarised electromagnetic fields at multiple positions within the unit cell \cite{Sollner2015, Mahmoodian2017, MurendranathPatil2022, Ostfeldt2022, Siampour2023, Germanis2025,Martin_comparison}. In addition, operation close to the slow-light region enhances the local density of optical states, strengthening light-matter interaction and enabling Purcell-enhanced emission into the guided mode \cite{Hughes2005,Gonzalez-Tudela2017}. Recent experiments using InAs quantum dots in glide-plane waveguides have exploited the slow-light regime to demonstrate strongly directional emission together with Purcell-enhanced radiative decay and waveguide-coupling efficiencies approaching unity \cite{Siampour2023,Germanis2025}.

Previous work on glide-plane waveguides has largely focused on emitters near the waveguide centre, where the guided mode provides strong and robust directionality \cite{Siampour2023,Germanis2025,Martin_comparison}. In that regime, however, the sign of the chiral coupling is effectively fixed by the emitter position because the local chirality varies only weakly with wavelength. Here we address the complementary and less explored regime of off-centre emitters. Numerical simulations predict that, close to the slow-light band, these positions can host chiral inversion points: fixed spatial locations where the handedness of the local guided field changes sign as the wavelength is tuned. This suggests a qualitatively different form of control, in which the directionality of a fixed emitter can be reversed spectrally without changing its position. In this work we test that prediction by Stark tuning a single Zeeman-split quantum-dot transition through the slow-light region of a glide-plane waveguide. We observe a sign reversal in the measured directional contrast and show that the behaviour is consistent with a reversal of the local chirality driven by a wavelength-dependent change in the relative phase of the local field components at an off-centre emitter position.

\section{DEVICE DESIGN}

We investigate the glide-plane photonic-crystal waveguide (GPW) device shown in Fig.~\ref{fig1}(a). The structure is fabricated from a suspended GaAs membrane containing self-assembled InAs/InGaAs quantum dots embedded in the intrinsic region of a vertical diode. The wafer consists of a Si-doped n-GaAs contact layer, a lower AlGaAs barrier region, the InAs quantum-dot layer with a thin GaAs cap, an upper AlGaAs barrier, and a GaAs / carbon-doped p-GaAs top-contact region. During nanofabrication, the membrane is released by selectively removing the underlying higher Al content AlGaAs sacrificial layer, yielding a suspended photonic-crystal platform with low optical loss. A schematic of the diode structure is shown in Fig.~\ref{fig1}(b).
The device follows the same basic design principles as earlier GPW platforms \cite{Sollner2015,Siampour2023,Germanis2025}, but incorporates improved mode adaptors to increase the usable collection efficiency in photoluminescence experiments. The photonic circuit consists of a central glide-plane section engineered to support slow-light propagation, followed by mode-adapting sections that convert the photonic-crystal mode into fast-light nanobeam waveguides. Shallow etch grating out-couplers at both ends of the device enable collection of waveguide-coupled photoluminescence, while electrical contacts to the p- and n-doped layers provide bias control across the diode. This architecture allows a fixed QD to be tuned spectrally across the slow-light band while its emission into the guided mode is monitored from the out-couplers.

\section{NUMERICAL SIMULATIONS}
\label{simulations}

\begin{figure}[t]
\includegraphics[width=0.5\textwidth]{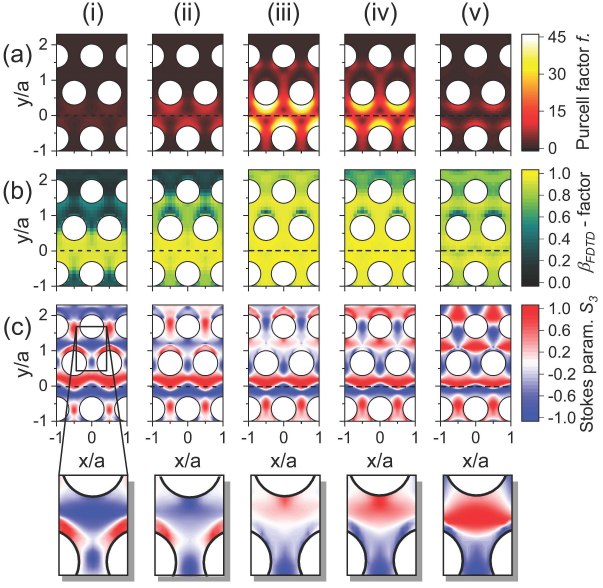}
\caption{Spatial dependence of simulated (a) Purcell factor, (b) waveguide coupling efficiency ($\beta_{\mathrm{FDTD}}$-factor), and (c) the local Stokes parameter $S_3$, with (i)--(v) corresponding to normalised frequencies as depicted in Fig.~\ref{fig1}(c). (i.e. $a/\lambda=0.2810, 0.2833, 0.2834, 0.2838, 0.2895$) The horizontal ($x$) and vertical axes ($y$) denote the longitudinal and lateral displacement respectively. Fields are calculated at the middle of the membrane, at the location of the quantum dots. While the chirality near the central region remains comparatively stable across the sampled wavelengths, off-centre regions exhibit a much stronger spectral dependence, including inversion of the sign of $S_3$ between (i)--(v). The dashed line in (a-c) marks the centre of the waveguide.}
\label{fig2}
\end{figure}
We start by presenting numerical simulations to predict the local electromagnetic environment experienced by an embedded QD and, from this, identify where inversion of directional emission should occur. Specifically, we determine the guided-mode dispersion and slow-light spectral region, evaluate the position-dependent coupling strength through the Purcell factor and waveguide $\beta$ factor, and calculate the local optical chirality through the Stokes parameter $S_3$. Together, these quantities predict the regions of the unit cell in which electrically tunable inversion of directional coupling is expected to be observable.

The simulated band structure and corresponding group index are shown in Fig.~\ref{fig1}(c). The lower guided band supports a pronounced slow-light region, where the large group index both enhances the local density of optical states and makes the spatial profile and polarization of the Bloch mode strongly wavelength dependent. This second feature is central to the present work. Whereas slow light is often invoked only for Purcell enhancement, here it also creates the spectral dispersion of the local polarization field required for a fixed emitter to cross a chiral inversion point under electrical tuning. Five representative spectral positions across this band, labelled (i)- (v), are highlighted in Fig.~\ref{fig1}(c) and are used in Fig.~\ref{fig1}(d-e) to illustrate how the directional response of an off-center circular dipole evolves with wavelength. Cases i and v denote wavelengths far detuned from the slow-light region; cases ii and iv mark the spectral position with near-unity chirality (but opposite sign) at the upper and lower branch of the dispersion curve, respectively; and case iii indicates the peak of the group index. A more detailed spatial dependence and the tunability of the parameter $S_3$ is presented in Fig.~\ref{figs3}, Appendix C. In previous work we have focused on the region with $y/a < 0.2$, i.e. the central part of the waveguide. Here $y$ is the lateral displacement from the centre of the waveguide and $a$ is the lattice constant (see Fig.~\ref{fig2} and Fig.~\ref{figs3}), see also the definition of $x,y,z$ axes in Fig.~\ref{fig1}(a). In this region, the chirality is large and does not vary significantly with wavelength ($\lambda$). However, a much richer range of behaviour is found for increasing values of $y/a$. In particular, for an emitter far displaced from the waveguide centre (i.e. $y/a>0.5$) and spectrally close to the slow-light region, we note that the polarisation of the local electric field profile varies rapidly with wavelength, thus introducing a drastic change in directionality over a small wavelength range. It is this inversion of the local handedness that we investigate in this paper. A full, detailed spatial and spectral variation in directionality is given in the Appendix~\ref{S3}.

A central requirement for observing off-centre chiral effects is that the QD couples sufficiently strongly to the guided mode at its location. Figure~\ref{fig2} therefore presents spatial maps of the Purcell factor $F_{\rm P}$, waveguide coupling efficiency ($\beta$), and local chirality ($S_3$) within a unit cell for five representative wavelengths as depicted in Fig.\ref{fig1}. Here $x$ is the coordinate along the waveguide axis and $y$ is the lateral displacement from the waveguide centre as shown in Fig.~\ref{fig1}(a).

The Purcell factor map in Fig.~\ref{fig2}(a) is calculated using guided-mode expansion (GME) implemented in \textit{Legume} \cite{Minkov2020} and is obtained from
\begin{equation}
F_{\rm P}(\bm{r})=\frac{3\pi c^2a}{\omega^2\sqrt{\epsilon_{\rm r}(\bm{r})}}\cdot\frac{1}{v_{\rm g}}\cdot\frac{|\hat{e}_k(\bm{r})\cdot\hat{n}|^2}{\int\epsilon_{\rm r}(\bm{r})|\hat{e}_k(\bm{r})|^2\rm{d}^3r}\:,
\label{Purcell}
\end{equation}
where $\hat{e}_k(\bm{r})$ is the Bloch mode profile, $\hat{n}$ is the dipole orientation, $v_{\rm g}$ is the group velocity, and $\epsilon_{\rm r}$ is the dielectric permittivity. As expected for photonic-crystal waveguides, $F_{\rm P}$ is strongly position dependent and increases substantially in the slow-light regime, reflecting the enhanced local density of optical states \cite{Mahmoodian2017,MurendranathPatil2022}.

We also evaluate the waveguide $\beta$ factor, which quantifies the fraction of emission coupled to the guided mode:
\begin{equation}
\beta=\frac{\Gamma_{\rm wvg}}{\Gamma_{\rm wvg}+\Gamma_{\rm free}+\gamma_{\rm nr}}\:.
\label{beta}
\end{equation}
where $\Gamma_{\rm wvg}$ and $\Gamma_{\rm free}$ are the decay rates into the waveguide mode and free space, respectively and $\gamma_{\rm nr}$ is the nonradiative decay rate.
The simulated $\beta_{\rm FDTD}$ shown in Fig.~\ref{fig2}(b) is obtained using \textit{Lumerical FDTD Solutions} \cite{Lumerical} and excludes non-radiative decay channels (i.e. it assumes $\gamma_{\rm nr}=0$). This approximation is not valid in the tunnelling-dominated regime at large reverse bias (see Section~\ref{slow light region}), but is appropriate elsewhere and is consistent with the observation that QDs can remain bright even when radiative rates are inhibited in photonic-crystal environments \cite{Kaniber2008}. Importantly, while efficient coupling outside the slow-light region is largely confined near the waveguide centre, within the slow-light region $\beta_{\rm FDTD}>0.9$ persists over a substantially larger area, including positions displaced far from the centre (e.g. $y/a>1$). This expanded high-$\beta$ region enables experimental access to off-centre QDs in previously unexplored regions of the unit cell.

Having established that slow light enables efficient waveguide coupling over a wide range of positions, we now analyse the local optical chirality of the guided mode. The key feature explored in this work is the existence of chiral inversion points: fixed locations within the waveguide unit cell for which the local optical chirality crosses zero and changes sign as the emission wavelength is tuned. We quantify the local optical chirality using the Stokes parameter $S_3$ of the in-plane electric-field components of the forward-propagating mode. In normalised form,
\begin{equation}
\label{eq:S3S0}
\begin{aligned}
\frac{S_3(\mathbf{r},\lambda)}{S_0(\mathbf{r},\lambda)}
&=\frac{2\,\mathrm{Im}\!\left\{E_x(\mathbf{r},\lambda)E_y^*(\mathbf{r},\lambda)\right\}}
{|E_x(\mathbf{r},\lambda)|^2+|E_y(\mathbf{r},\lambda)|^2} \\
&=\frac{2|E_x||E_y|}{|E_x|^2+|E_y|^2}\,\sin\!\big(\Delta\phi(\mathbf{r},\lambda)\big),
\end{aligned}
\end{equation}
where $\Delta\phi(\mathbf{r},\lambda)=\phi_x-\phi_y$ is the relative phase between the field components.

Equation~\eqref{eq:S3S0} makes the inversion mechanism explicit. The prefactor is positive and cannot change sign, so variations in field intensity can only reduce $|S_3/S_0|$ and, in the extreme case, drive it to zero if one component vanishes. A sign inversion of $S_3/S_0$, however, requires $\sin(\Delta\phi)$ to change sign, i.e. a wavelength-dependent change of the local relative phase corresponding to a reversal of the handedness of the polarisation ellipse.

The chirality maps in Fig.~\ref{fig2}(c) show that $S_3$ can change significantly with wavelength at off-centre locations, consistent with the enhanced sensitivity introduced by slow light \cite{Siampour2023}. The simulated wavelength dependence of the local chirality is summarised in Fig.~\ref{fig2}(c), plotted as a function of lateral displacement $y/a$ (with $a=259$~nm). Near the central region previously exploited for high directionality (typically $y/a\sim 0.2$), the chirality remains large and varies only weakly with wavelength. In contrast, for more off-centre positions the local polarisation becomes highly dispersive near the slow-light band, where the mode profile changes most rapidly with wavelength, and $S_3/S_0$ can switch rapidly from $\sim -1$ to $\sim +1$ over a narrow wavelength interval. These off-centre regions therefore host chiral inversion points that can be traversed experimentally by tuning the QD emission wavelength while the emitter position remains fixed.

\begin{figure*}[t]
\includegraphics[width=0.8\textwidth]{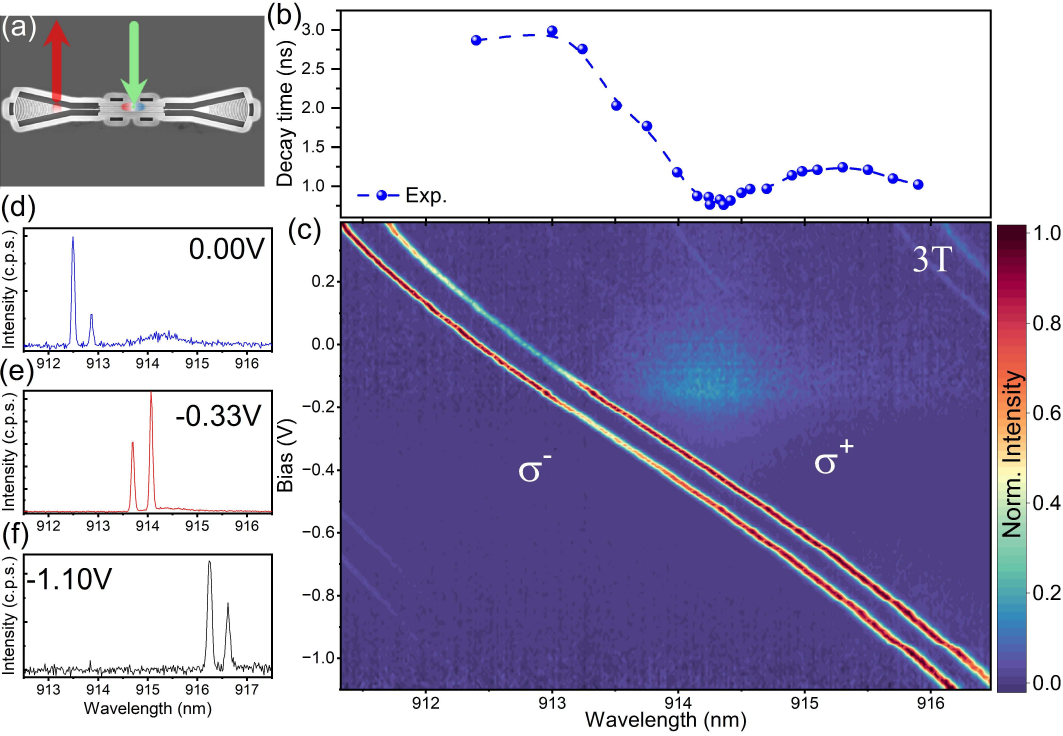}
\caption{(a) Schematic of the device and excitation scheme; green and red arrows indicate excitation and collection positions. (b) QD exciton decay times as a function of wavelength over the slow-light bandwidth. (c) Photoluminescence map of non-degenerate circularly polarised excitons at 3~T (Faraday geometry), as a function of externally applied bias. (d)-(f). Characteristic photoluminescence spectra recorded upon the application of external bias at $0.00$~V, $-0.33$~V, and $-1.10$~V, respectively.}
\label{fig3_4}
\end{figure*}

In Section~\ref{Experimental results} we present experimental data from a QD whose directional contrast exhibits a sign inversion as its emission is tuned across the slow-light region. Since QDs are randomly positioned, the emitter location must be inferred by comparing experiment to simulations. The wavelength dependence of the measured QD lifetime (Fig.~\ref{fig3_4}(b)) and the requirement of measurable guided-mode intensity constrain the emitter to regions that exhibit inhibited emission outside the slow-light band ($F_{\rm P}<1$) while maintaining a sufficiently large $\beta$ factor. Figure~\ref{figs3}(a) in Appendix C, indicates a set of candidate positions (yellow markers) that satisfy these constraints, and highlights the best-estimate position (black marker). At such off-centre positions, the simulated evolution of $S_3$ across the slow-light bandwidth predicts a sign change of the local chirality at fixed $\mathbf r$, which qualitatively accounts for the experimentally observed wavelength-dependent evolution and inversion of the directional emission contrast that we now present.

\section{EXPERIMENTAL RESULTS}
\label{Experimental results}
\subsection{Identification of the slow-light spectral region}
\label{slow light region}

We first identify the slow-light spectral region of the device using time-resolved photoluminescence measurements, Fig.\ref{fig3_4} (a) shows the excitation scheme used for the measurements. The exciton lifetime is extracted as a function of emission wavelength while the QD transition is electrically tuned across the relevant spectral range.

Figure~\ref{fig3_4}(b) shows the measured exciton decay time versus wavelength and representative decay curves are shown in Fig.~\ref{fig_9}. The lifetime reaches a minimum of $\sim 760$~ps at 914.3~nm, which defines the centre of the experimentally accessible slow-light region. This assignment is independently supported by the maximum in the out-coupled intensity shown in Fig.~\ref{fig7}(b). For wavelengths $\lambda<913$~nm, a long decay time of $\sim 3$~ns is observed, which is significantly longer than the typical exciton lifetime ($\sim 1.0$--1.3~ns) measured from this sample outside photonic devices. This indicates inhibited spontaneous emission due to the reduced local density of optical states in the photonic band-gap region \cite{Koenderink2006,Kaniber2007,Kaniber2008,Wang2011}. Taken together with the PL intensity detected at the out-coupler, these observations constrain the QD position to regions displaced from the waveguide centre, while excluding locations so far from the centre that the $\beta$ factor is too low for the QD emission to couple efficiently into the guided modes and be collected.

For wavelengths $\lambda>914.3$~nm the lifetime initially increases and then decreases again for $\lambda\gtrsim 915.3$~nm. The reduction at longer wavelengths is attributed to increased carrier tunnelling rates induced by the applied electric field, which introduces non-radiative decay channels. In this tunnelling-dominated regime the approximation $\gamma_{\rm nr}=0$ in Eq.~\ref{beta} is no longer valid. 

\subsection{Traversing a chiral inversion point by Stark tuning}

Figure~\ref{fig3_4}(c) presents bias-dependent normalised photoluminescence (PL) spectra of the QD under study at $B=3$~T. The magnetic field is applied along the growth direction (Faraday geometry), splitting the exciton into two non-degenerate spin states with opposite circular polarisations via the Zeeman effect. A 3~T magnetic field induces a Zeeman splitting of approximately 0.53~meV between the $\sigma^+$ and $\sigma^-$ transitions. By varying the external bias, the emission energy is tuned continuously via the quantum-confined Stark effect. Throughout these measurements the excitation conditions are kept constant. The excitation laser is linearly polarised and has an energy well detuned above the exciton transition in the QD's wetting layer, thereby excluding spin-selective excitation and optical spin pumping effects \cite{Germanis2025}. Directional emission is quantified by comparing the intensities of the $\sigma^+$ and $\sigma^-$ Zeeman transitions collected from the same out-coupler under fixed excitation conditions. The observed intensity evolution therefore reflects changes in the coupling between the QD emission and the waveguide mode as the emission wavelength is tuned through the slow-light region identified above.

\begin{figure}[htbp]
\includegraphics[width=8.5cm]{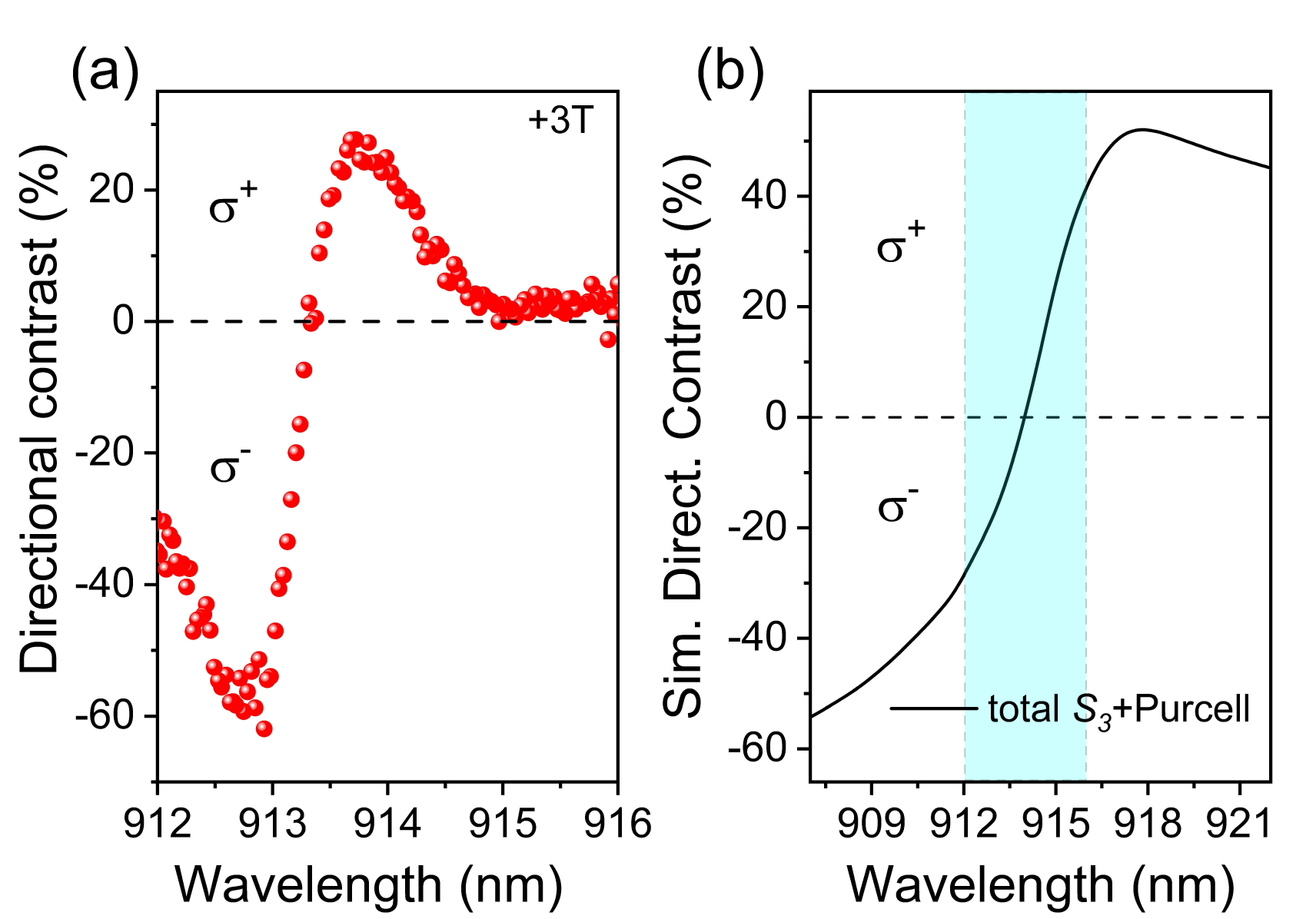}
\caption{(a) Directional contrast recorded from the left out-coupler using the PL map in Fig~\ref{fig3_4}(c). (b) Simulated directional contrast for the candidate QD position, including contributions from the intrinsic chirality $S_3$ and Purcell enhancement. The cyan shaded region denotes the spectral bandwidth corresponding to the experimental data shown in panel (a).}
\label{fig5}
\end{figure}

The main panel of Fig.~\ref{fig3_4} (c) and selected spectra (d-f) show the tuning of the two circularly polarised Zeeman transitions across the spectral range spanning the slow-light region. A pronounced and opposite evolution of the intensities of the two transitions is observed: one spin state increases in collected intensity while the other decreases under identical excitation conditions, and the trend reverses at higher biases. The directional contrast $D$ is extracted from the integrated intensities of the two Zeeman components ($\sigma_+$ and $\sigma_-$) measured from the same out-coupler,
\begin{equation}
D = \frac{I_{\sigma_+} - I_{\sigma_-}}{I_{\sigma_+} + I_{\sigma_-}}\:,
\end{equation}
where $I_{\sigma_+}$ and $I_{\sigma_-}$ are the fitted peak areas of the $\sigma_{+/-}$ exciton lines. Figure~\ref{fig5}(a) shows the directional contrast deduced from Fig.~\ref{fig3_4}(c) as the exciton emission wavelength is tuned across the slow-light band. The directional contrast varies strongly with wavelength and changes sign at 913.3~nm.

Experimentally, we measure the directional contrast $D$, not the local chirality $S_3$ directly. The interpretation therefore relies on comparing the data with simulated candidate emitter positions and on assessing alternative mechanisms that could also distort $D$. The most obvious alternative is spectral shaping, in which the Zeeman-split transitions experience slightly different Purcell enhancements as they pass through the slow-light region. Such effects can modify the apparent directionality in resonant photonic systems \cite{Rao2025,Martin2025}. For the present device, however, the calculated contribution from spectral shaping is small and does not reproduce the observed sign reversal (Appendix~\ref{appendix}). We therefore conclude that the dominant origin of the measured inversion of $D$ is the wavelength dependence of the intrinsic local chirality at a fixed position of an off-center emitter.

Figure~\ref{fig5}(b) shows the simulated wavelength dependence of the directional contrast for the candidate QD position obtained from the FDTD and GME calculations described in Section~\ref{simulations}. The simulation reproduces the main qualitative features of the experiment, namely the strong spectral dependence of $D$ and its sign reversal within the slow-light region. We note that the measured slow-light feature is narrower than in the idealised model, most likely due to fabrication-induced deviations from the nominal geometry, as discussed in Appendix~\ref{slow light band}. Nevertheless, the combined lifetime, intensity, and directionality data are all consistent with a fixed off-centre emitter whose local chirality changes sign as the emission wavelength is tuned. The key result is therefore not simply wavelength-dependent directionality, but electrically induced reversal of the chiral coupling sign for a fixed emitter in a nanophotonic waveguide.

\section{Conclusion}
We have demonstrated electrically tunable chiral light--matter interaction for a single InAs/InGaAs quantum dot embedded in a glide-plane photonic-crystal waveguide operating in the slow-light regime. Time-resolved photoluminescence identifies the slow-light centre at 914.3~nm, and Stark tuning of the QD emission across the slow-light band reveals a strong wavelength dependence of the directional emission, including a sign inversion of the directional contrast.

By comparing the experimental data to numerical simulations, we infer that the QD is located away from the waveguide centre, beyond the first row of holes, where the local optical chirality varies strongly with wavelength. The measured sign inversion is therefore explained by the intrinsic spatial and spectral dependence of the local chirality, while spectral shaping due to differential Purcell enhancement of the Zeeman components provides only a minor correction. These results show that chiral coupling in glide-plane waveguides is not solely fixed by emitter position, but can instead be actively controlled in situ via electrical tuning of the emission wavelength.

This additional degree of control opens a route toward reconfigurable chiral quantum photonic devices, in which the emission direction of a fixed emitter can be selected dynamically after fabrication. More broadly, electrically accessing chiral inversion points provides a possible mechanism for tuning the effective emitter--waveguide coupling relevant to nonreciprocal photon transport and spin--photon interfaces \cite{Lodahl2017,PRXQuantum.6.020101}. In the longer term, combining such control and improved spatial selectivity of the emitters could enable active single-photon routing and switching functionalities, as well as tunable multi-emitter chiral networks implemented within a scalable waveguide-QED architecture \cite{Yan2011,Cheng2016,Li2018,Poudyal2020,Hallacy2025Dimers,switch_review}.


\begin{acknowledgments}
This work was funded by the Engineering and Physical Sciences Research Council (EPSRC) UK Programme Grant EP/V026496/1.
\end{acknowledgments}


\appendix

\section{Simulation vs Experimental Slow-Light Band}
\label{slow light band}

\begin{figure}[htbp]
\includegraphics[width=8.5cm]{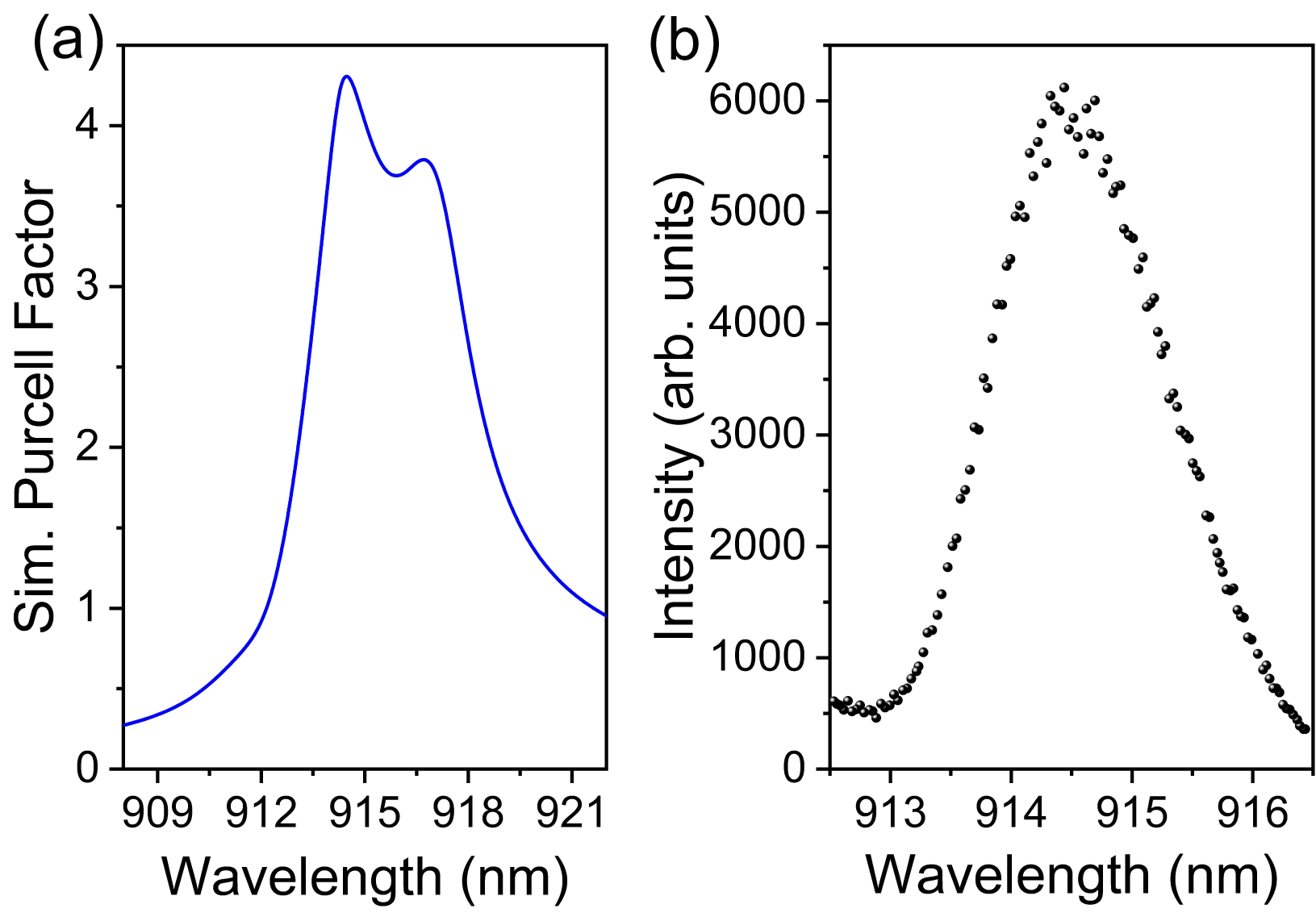}
\caption{(a) Simulated wavelength dependence of the Purcell factor at the candidate QD position. (b) Integrated QD intensity collected from the out-coupler as a function of wavelength (i.e. the average wavelength of the $\sigma_+$/$\sigma_-$ components) measured under the same conditions as Fig.~\ref{fig3_4}(b).}
\label{fig7}
\end{figure}

Figure~\ref{fig7}(a) shows the simulated wavelength dependence of the Purcell factor evaluated at the candidate QD position identified in the main text. The Purcell factor follows the dispersion of the local optical density of states, and exhibits two maxima within the engineered slow-light region, reflecting the double-peak structure of the group index shown in Fig.~\ref{fig1}(b).

The calculated Purcell factor can be compared to the spectrally integrated QD intensity as a function of the mean wavelength of the $\sigma_{+}/\sigma_{-}$ Zeeman transitions in Fig.~\ref{fig3_4}(b). The results are shown in Fig.~\ref{fig7}(b). If we assume that there is no non-radiative decay, the total integrated intensity should not vary with the radiative lifetime, as only one photon can be emitted per laser pulse. However, in our experiment the QD PL intensity measured from an out-coupler, or the detected photon number, depends on the fraction of decay channels that are radiative and are coupled into the waveguide mode (i.e. the $\beta$ factor defined in Eq.~\ref{beta}). Hence the integrated intensity $I(\lambda)$ scales approximately as

\begin{equation}
\label{Intensity}
I(\lambda)\propto\beta_{\rm{FDTD}}(\lambda)\cdot\left(1-\gamma_{\rm{nr}}/\Gamma_{\rm{exp}}\right)\:.
\end{equation}

Here $\beta_{\rm{FDTD}}$ is the radiative coupling efficiency of the emitter into the waveguide mode simulated by FDTD as in eqn.\ref{betaFDTD}, $\Gamma_{\rm{exp}}$ is the experimentally measured total decay rate of an exciton state and $\gamma_{\rm{nr}}$ denotes the non-radiative decay rate which is treated as a free parameter in our analysis. To estimate the variation in integrated intensity and directional contrast, we include a non-radiative decay rate of $\gamma_{\rm nr}\sim 0.17\,{\rm ns}^{-1}$ in our model. The actual value of $\gamma_{\rm nr}$ is unknown, but the measured lifetime of $\sim 3$\,ns outside the slow-light region sets an upper bound of $\sim 0.33\,{\rm ns}^{-1}$. The value chosen gives a reasonable fit to the data, and has little effect in the slow-light region where the radiative rate is strongly enhanced. This shows that the waveguide-coupled intensity does peak in the slow-light region, where the simulated $\beta_{\rm{FDTD}}$ is maximized.

On comparing Figs.~\ref{fig7}(a) and (b), we immediately notice that the experimental peak is much narrower than the simulations would suggest. Furthermore, the maximum value of $F_{\rm P}$ deduced from the minimum in the lifetime in Fig.~\ref{fig3_4} and Fig.~\ref{fig_9} is only $\sim 1.3$ compared to $>4$ in the simulation. These observations are commonly reported in photonic crystal waveguides \cite{Scarpelli2019,Germanis2025,Mann2015,Siampour2023}, where the physical origin is primarily attributed to fabrication-induced disorder, which reduces the effective group index and perturbs the slow-light performance of the photonic device \cite{Hughes2005,Patterson2009_coherent,Patterson2009_incoherent}. As a result, the measured Purcell factor is typically lower than that predicted for an ideal structure and exhibits a narrower slow-light bandwidth.
Additionally, close inspection of the spectrum at 0.00 V Fig.~\ref{fig3_4} (d) and the photoluminescence map from 0.00 V to -0.12 V Fig.~\ref{fig3_4} (c) reveals a weak, broad emission peak centred at 914.3 nm, which may be associated with an enhanced phonon sideband on the low-energy Stokes side of the QD emission. Its appearance is consistent with a tuning range that spans the slow-light region.

\section{Directional Contrast and Spectral Shaping Model}
\label{appendix}
The directional contrast is simulated by modelling the two Zeeman components as circularly polarised dipoles. The detected PL intensity of a given Zeeman component under pulsed excitation could be written as the product of two factors $\eta_{\pm}(\bm{r}, \lambda)$ and $\beta(\bm{r}, \lambda)$ -- the former accounts for the fraction of photons that propagate towards the target direction, while the latter quantifies the overall coupling efficiency into guided modes. For a circularly polarised dipole at position $\mathbf{r}$, the probability of directional coupling is determined by the local Stokes parameter $S_3$ such that

\begin{equation}
\eta_{\pm}(\bm{r}, \lambda)=\frac{I_L}{I_L+I_R}=\frac{1\pm S_3(\bm{r}, \lambda)}{2}\:,
\end{equation}
where $I_{L/R}$ denotes the collected PL intensity at the left/right out-coupler. Substituting $\beta(\bm{r}, \lambda)$ to Eqn.\ref{beta} then yields

\begin{equation}
\label{intermediate}
I_{\pm}(\bm{r}, \lambda)\propto \frac{1\pm S_3(\bm{r}, \lambda)}{2}\cdot\frac{\Gamma_{\rm wvg}}{\Gamma_{\rm wvg}+\Gamma_{\rm free}+\gamma_{\rm nr}}\:.
\end{equation}

\begin{figure}[t]
\includegraphics[width=8.5cm]{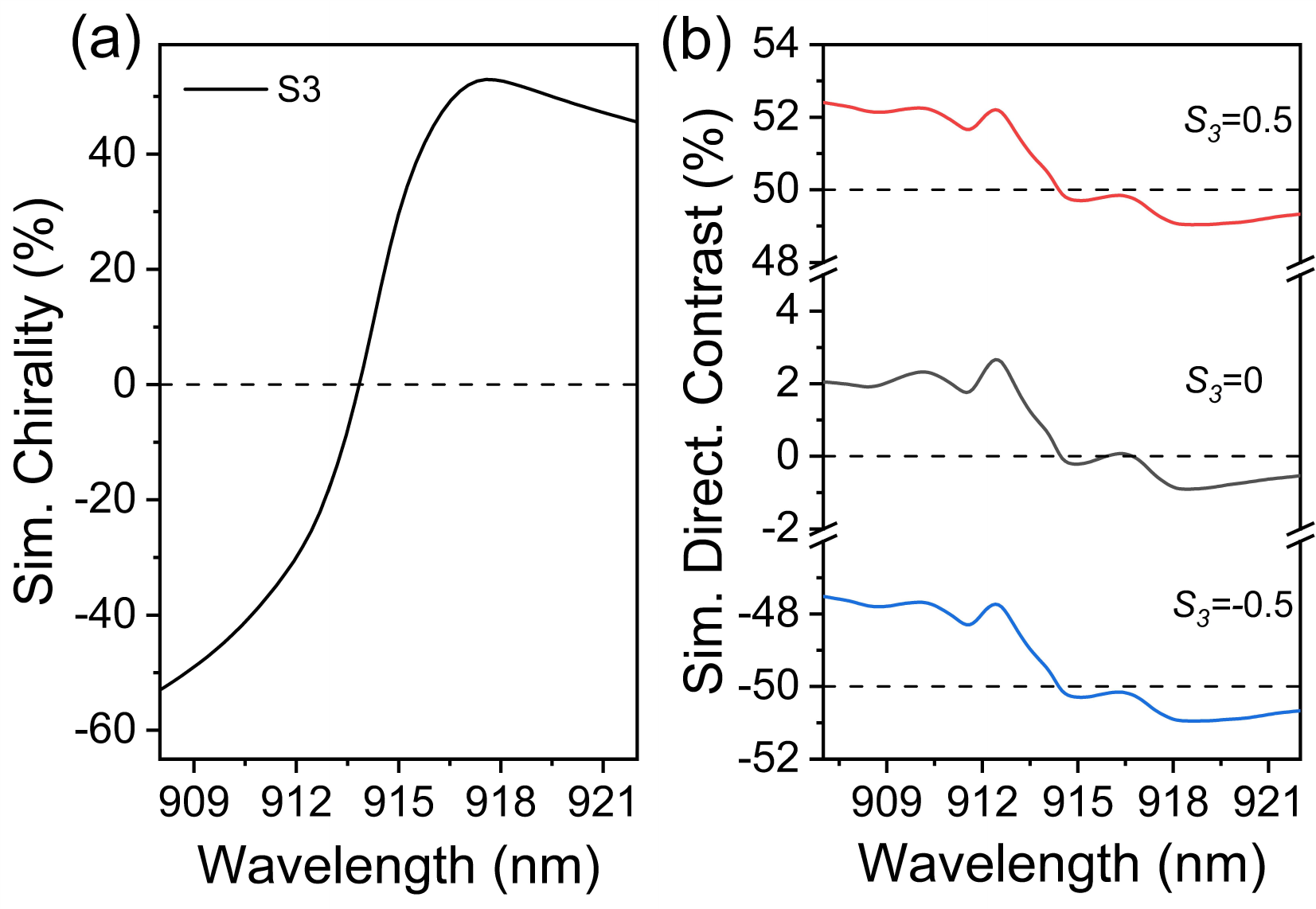}
\caption{(a) Simulated wavelength dependence of the directional contrast deduced from the local Stokes parameter $S_3$ at the candidate QD position. (b) Calculated directional contrast across the slow-light region for fixed intrinsic chirality values $S_3=-0.5$, 0, and $0.5$. A Zeeman splitting of $\Delta E=0.53$~meV is assumed.}
\label{fig6}
\end{figure}

In FDTD simulation, the coupling efficiency $\beta_{\rm{FDTD}}$ accounts only for the radiative decay channels and is written as

\begin{equation}
\label{betaFDTD}
\beta_{\rm{FDTD}}=\frac{\Gamma_{\rm wvg}}{\Gamma_{\rm wvg}+\Gamma_{\rm free}}\:.
\end{equation}
Rearranging this equation and substituting the radiative decay rate $\Gamma_{\rm wvg}$ into the simulated Purcell factor $F_{\rm P}\Gamma_0$, where $\Gamma_0$ is the radiative decay rate in a homogeneous medium (i.e. bulk GaAs), gives
\begin{equation}
\Gamma_{\rm wvg}+\Gamma_{\rm free}=\frac{F_{\rm P}\Gamma_0}{\beta_{\rm{FDTD}}}\:.
\end{equation}
Finally, by combining the above relations, the expression of Eqn.\ref{intermediate} becomes

\begin{equation}
\label{Zeeman intensities}
I_{\pm}(\bm{r}, \lambda)\propto \frac{1\pm S_3(\bm{r}, \lambda)}{2}\cdot\frac{F_{\rm{P}}(\bm{r}, \lambda)\,\Gamma_0}{F_{\rm{P}}(\bm{r}, \lambda)\,\Gamma_0/\beta_{\rm{FDTD}}(\bm{r}, \lambda)+\gamma_{\rm{nr}}}\:,
\end{equation}

where $I_{\pm}$ is the intensity of the $\sigma_{\pm}$ transition collected from the left out-coupler. Figure~\ref{fig6}(a) shows the simulated wavelength dependence of the local Stokes parameter $S_3(\lambda)$ at the candidate QD location identified in the main text. The smooth evolution of $S_3$ across the engineered slow-light band provides a clear and direct mechanism for the experimentally observed inversion of the directional contrast shown in Fig.~\ref{fig5}(a).
  
The second factor in Eq.~\ref{Zeeman intensities} is included to account for the spectral shaping effect that arises from the different Purcell enhancements of the two Zeeman components when they are tuned through the slow-light region. To evaluate the magnitude of this effect, we calculate the directional contrast for several fixed values of the intrinsic chirality $S_3$. The results are shown in Fig.~\ref{fig6}(b), where $S_3$ is set to constant values ($-0.5$, 0, $+0.5$) throughout the entire range while the Purcell factor and the simulated $\beta_{\rm{FDTD}}$ factor retain their full wavelength dependence.

The curves can be interpreted intuitively by noting that the spectral separation of the Zeeman components is relatively small (0.53~meV) compared to the width of the slow-light region. Within a first-order approximation, the difference in the Purcell enhancements of the Zeeman components is proportional to the slope of radiative coupling, so that the imbalance between the two Zeeman components, and hence the change in the directionality, scales as
\begin{equation}
\label{slope}
\Delta D \propto \Delta\lambda\frac{\mathrm{d}(F_P\cdot\beta_{\rm{FDTD}})}{\mathrm{d}\lambda}
\end{equation}
where $\Delta \lambda$ is the Zeeman splitting. Consequently, when both Zeeman components lie on the same side of the Purcell factor peak, the contribution of the slow-light effect retains the same sign. In contrast, when the two components sweep across the slow-light centre, an opposite contribution is expected as the slope of both Purcell factor and simulated $\beta_{\rm{FDTD}}$ factor changes sign. Because a steeper slope is predicted for wavelengths blue-shifted from the slow-light centre, a larger deviation from the baseline is captured at shorter wavelengths.

\begin{figure}[htbp]
\includegraphics[width=8.5cm]{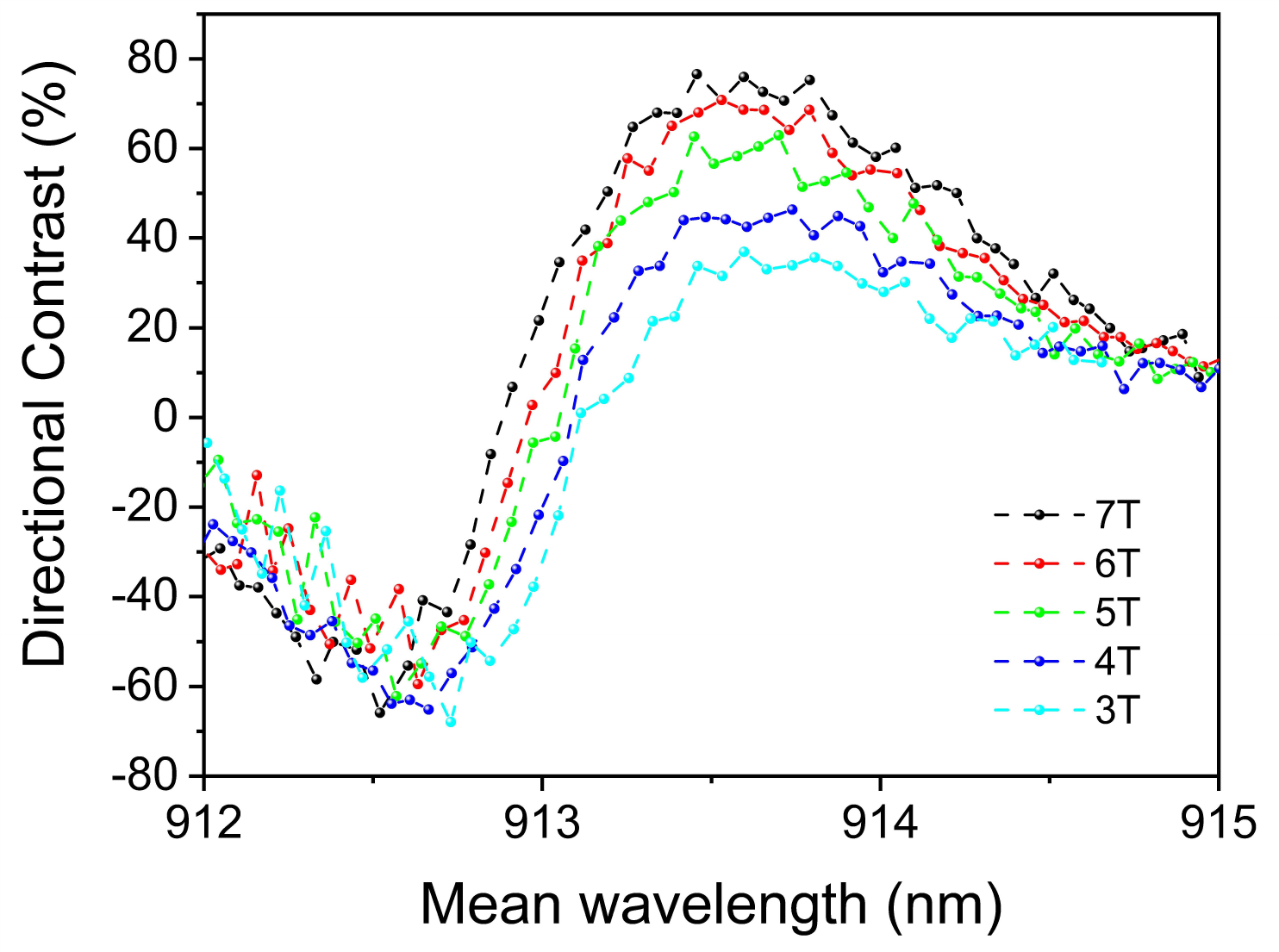}
\caption{Directional contrast dependence for various values of the applied magnetic field in Faraday configuration up to 7~T.}
\label{fig8}
\end{figure}

For fixed intrinsic chirality, the spectral shaping effect modifies the directional contrast by no more than $\sim 3\%$ across the slow-light region as shown in Fig.~\ref{fig6}(b). This indicates that asymmetric Purcell enhancement introduces only a minor modulation of the directionality. This is apparent in the experimental data shown in Fig.~\ref{fig8}, where $D$ is measured at different values of magnetic field $B$ from 3--7~T. The magnitude of $D$ varies with $\Delta \lambda$ according to Eq.~\ref{slope}. Although a larger contribution from spectral shaping is observed due to the narrowed bandwidth of the slow-light section compared with our simulation, the sign inversion remains largely unaffected. Therefore, we conclude that the experimentally observed sign inversion of the directional contrast arises primarily from the wavelength dependence of the intrinsic chirality of the photonic device.

\begin{figure*}[t]
\includegraphics[width=\textwidth]{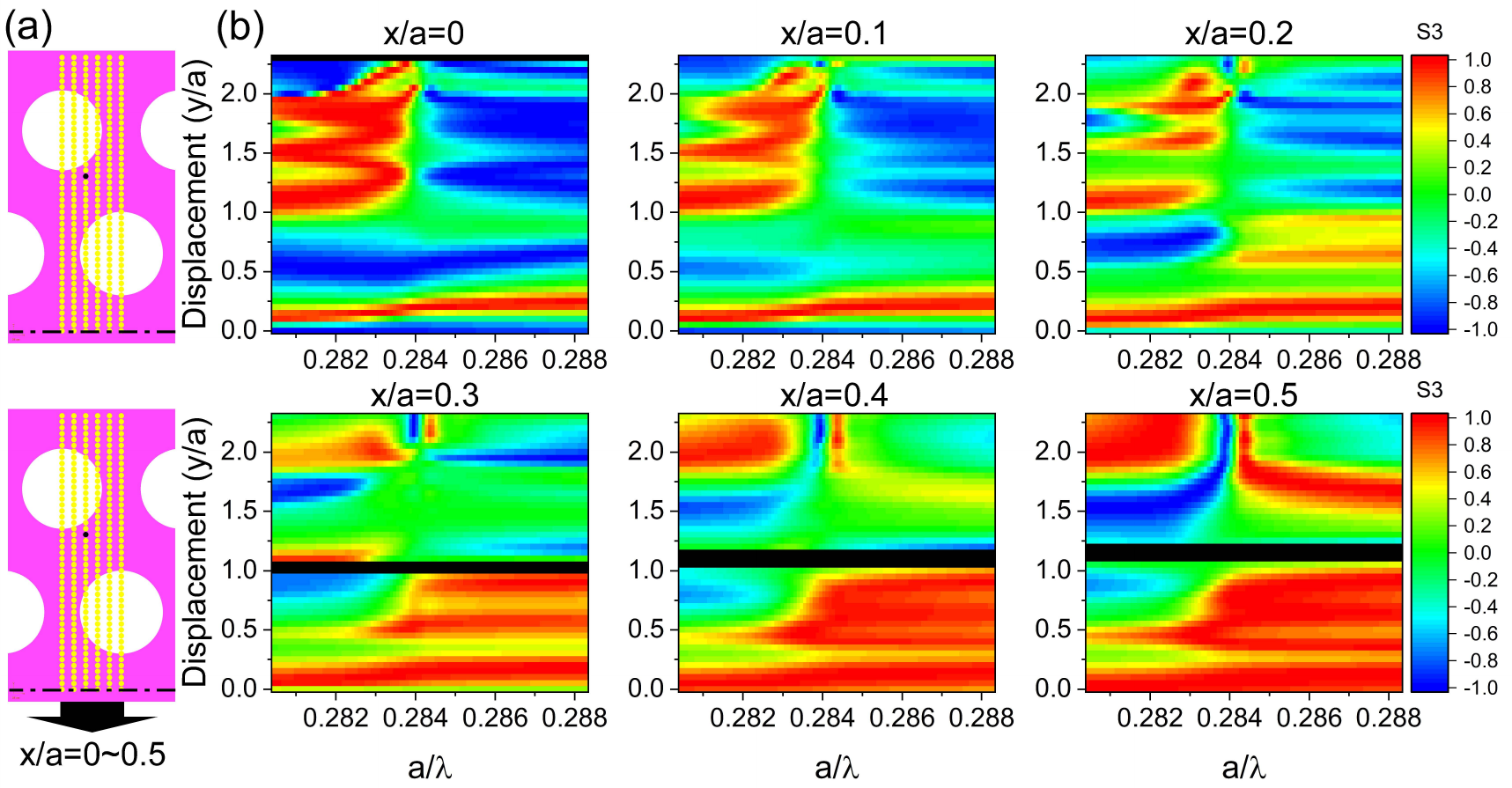}
\caption{(a) A detailed CAD view of the active region of our GPW device showing simulated candidate emitter positions (yellow markers), with the dash dot line indicating the centre of the waveguide (i.e. $y/a=0$). The coordinates are defined in accordance with Figure~\ref{fig2}, where $x$ and $y$ denote the longitudinal and lateral displacement respectively. The best candidate position for the QD studied experimentally is highlighted with a black marker. (b) Spatial and spectral dependence of simulated local Stokes parameter $S_3$ for six longitudinal slices within a single unit cell, taken at $x/a=0, 0.1, 0.2, 0.3, 0.4, 0.5$. Each panel shows the evolution of $S_3$ as a function of normalised wavelength $a/\lambda$ (horizontal axis) and lateral displacement $y/a$ (vertical axis). The chirality response is shaded in black for positions where the relative intensity of the electric field of the Bloch mode is too low (i.e. $|E(\bm{r})|^2<0.05|E_{max}|^2$).}
\label{figs3}
\end{figure*}

\section{Spatial and Spectral Dependence of Chirality}
\label{S3}

Figure~\ref{figs3} shows the simulated wavelength-dependence of local chirality at six longitudinal positions within a single unit cell. Each slice displays the lateral variation in local Stokes parameter $S_3$ for a fixed longitudinal offset of $x/a$ in units of $y/a$. A detailed CAD view and schematic of simulated candidate emitter positions is depicted in Fig.~\ref{figs3}(a). Spatially speaking, when an emitter is placed close to the waveguide centre (i.e. $y/a \sim 0.2$), the chirality is large and does not vary significantly with wavelength ($\lambda$). This provides robust and reliable chiral points for uni-directional coupling. Spectrally speaking, at a short or long wavelength far detuned from the slow-light region, the waveguide mode is weakly dispersive, and the directionality persists over a broad spectral window regardless of the emitter location. On the contrary, for wavelengths close to the slow-light centre (i.e. $a/\lambda = 0.2834$) where the mode profile experiences the most drastic change against wavelength, the directionality rapidly switches handedness, varying from $\sim -1$ to $\sim +1$ over a small wavelength range. These results highlight the significance of engineering a single-mode waveguide with broadband slow-light, as it aids with the variation in the relative amplitude and phase between transverse and longitudinal electric field components, giving rise to much sharper spectral response of chirality.

Upon comparing the six longitudinal slices, we notice that the precise spectral position of the sign-flip behaviour and its wavelength sensitivity also varies with $x/a$. Moreover, the handedness of the guided mode profile is not necessarily invariant for a fixed lateral displacement, indicating that the relative phase of the Bloch mode experiences a $\pi$ phase shift (e.g. from $+\pi/2$ to $-\pi/2$) along the propagation direction within one period. Consequently, emitters placed close to the waveguide centre exhibit relatively stable chirality but lack tunability, while those off-centred ones enable spectral tuning of emission direction at the cost of weaker coupling strength and increased sensitivity to disorder.

\section{Lifetime Measurements}

\begin{figure}[hbp]
\includegraphics[width=8.5cm]{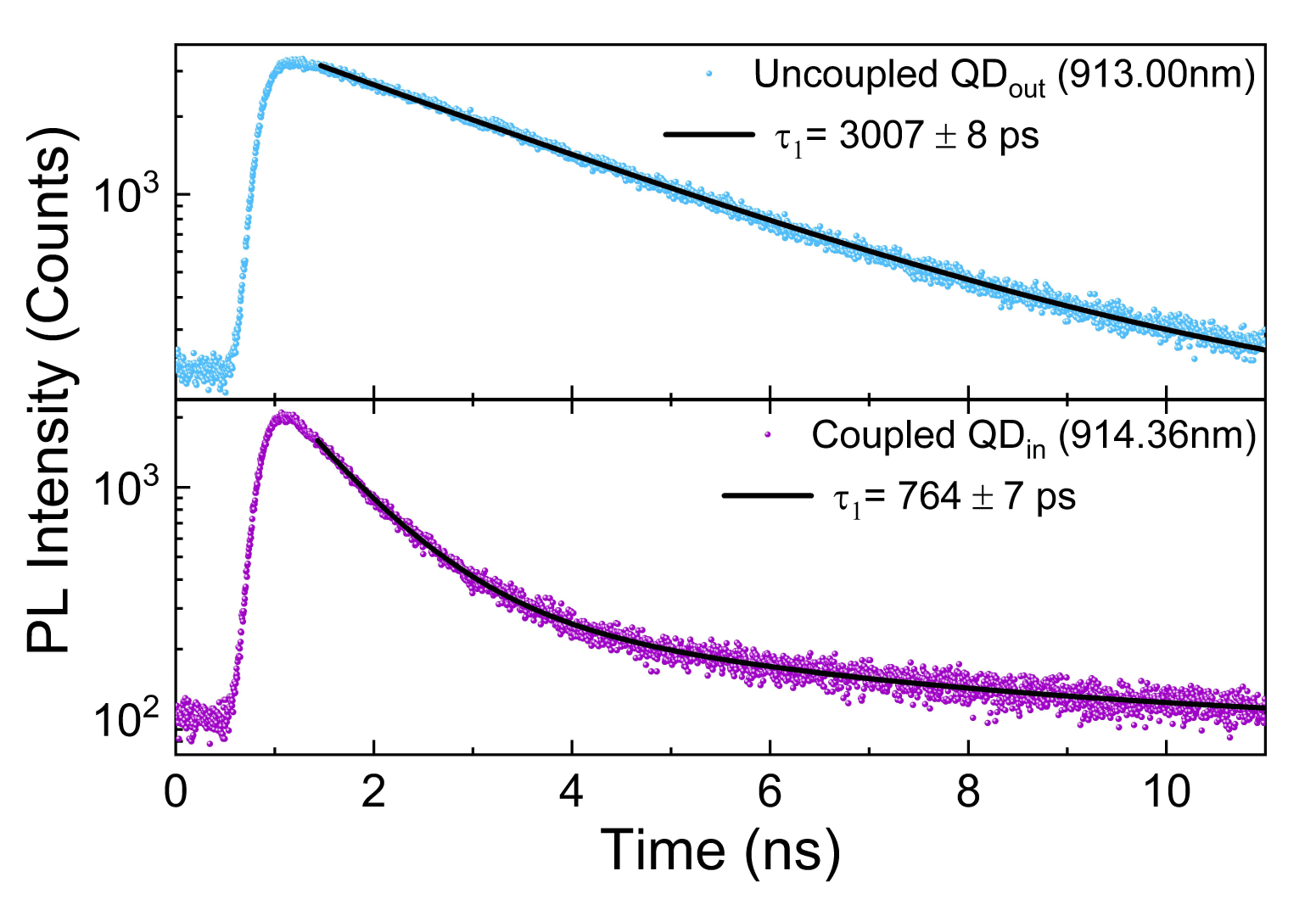}
\caption{Representative exciton decay curves recorded (\textit{Top panel}) far from the slow-light region, slow decay, and (\textit{Bottom panel}) at the centre of the slow-light region, the fastest decay.}
\label{fig_9}
\end{figure}

A representative decay curve of the exciton state uncoupled from the engineered photonic environment (i.e., outside the slow-light band) is shown in the top panel of Fig.~\ref{fig_9} and is characterised by a recombination time of approximately 3 ns. In contrast, a significantly faster recombination time, nearly four times faster, is extracted from the decay curve in the bottom panel, recorded when the exciton emission is coupled to the centre of the slow-light band. This behaviour indicates an effective enhancement of the radiative decay rate shown in Fig.~\ref{fig3_4} (b), due to the Purcell effect, as discussed in the main text. Outside the slow-light region, the decays are well described by single-exponential fits, whereas within the slow-light band bi-exponential behaviour is observed, consistent with modified recombination dynamics in the structured photonic environment \cite{Favero2005,Johansen2008,Smith2005}.

\section{Methods}
\label{Sample}

The sample is mounted in a helium bath cryostat. Optical excitation is performed using a pulsed laser (80~MHz repetition rate) at 870~nm, focused onto the central region of the waveguide. Photoluminescence is collected from a single out-coupler and directed to a spectrometer. Time-resolved measurements are performed using time-correlated single-photon counting. More details of the QD sample and optical setup are given in the Supplementary Information of Ref.~\cite{Germanis2025}.

\bibliographystyle{apsrev4-2}
\bibliography{Biblioo}

\end{document}